\documentstyle[aps]{revtex}
\begin{document}
\draft
%\tighten

\title{Tests and applications of self-consistent cranking 
in the interacting boson model}
\author{Serdar Kuyucak\thanks{E-mail: sek105@rsphysse.anu.edu.au}}
\address{Department of Theoretical Physics, Research School of Physical
	Sciences, \\
	Australian National University, Canberra, ACT 0200, Australia}
\author{Michiaki Sugita}
\address{Japan Atomic Energy Research Institute, Tokai, Ibaraki 
319-11, Japan}

\date{\today}
\maketitle

\begin{abstract}
The self-consistent cranking method is tested by comparing the 
cranking calculations in the interacting boson model with the exact 
results obtained from the SU(3) and O(6) dynamical symmetries and from 
numerical diagonalization.  The method is used to study the spin 
dependence of shape variables in the $sd$ and $sdg$ boson models.  
When realistic sets of parameters are used, both models lead to 
similar results: axial shape is retained with increasing cranking 
frequency while fluctuations in the shape variable $\gamma$ are 
slightly reduced.
\end{abstract}


\section{Introduction}
Self-consistent cranking (SCC) is one of the most popular methods to 
study collective rotations in nuclear many-body systems \cite{ring}.  
Within the Hartree-Fock framework, it has provided important insights 
into the backbending phenomena, and currently, it is being used 
actively in studies of superdeformed nuclei~\cite{bak95}.  In stark 
contrast, it has been rarely used in the interacting boson model (IBM) 
\cite{iac87}.  In fact, there is only one application of SCC to IBM 
where a study of moment of inertia (MOI) is carried out~\cite{sch85}.  
In the early stages of the model, an obvious reason for this neglect 
was the availability of exact results (either via dynamical symmetries 
or numerical diagonalization), which left little room for development 
of approximate methods such as SCC. In later extensions of the IBM 
(e.g.~$sdg$-IBM), dynamical symmetries were not very useful, and 
numerical diagonalization could not be carried out due to large basis 
space, hence alternative methods were needed.  At around the same 
time, however, an exact angular momentum projection technique was 
developed for axially symmetric boson systems ($1/N$ expansion 
\cite{kuy88}), which provided analytical solutions for general 
Hamiltonians in large basis spaces, filling the gap left by the other 
methods.  Although the assumption of axial symmetry is reasonable for 
most deformed nuclei in their ground states, triaxial effects are 
known to play a role, especially at high-spins \cite{abe90}.  
Therefore a study of the evolution of shapes in the IBM would be 
useful in order to investigate such questions as the effects of 
triaxiality in nuclear spectra, ways of incorporating it in the IBM 
Hamiltonian, and whether an improved description of spectra can be 
obtained if triaxial effects are taken into account.

In this paper, we first present tests of the SCC by comparing the 
cranking calculations with the exact results obtained in the SU(3) and 
O(6) limits, as well as from diagonalization of more realistic 
Hamiltonians in the $sd$-IBM. The SCC method is then used in a study 
of triaxiality in the IBM both in the $sd$ and $sdg$-boson versions of 
the model.

\section{Cranking formalism in the IBM}
The SCC in the IBM was formulated in Ref.~\cite{sch85} to which we 
refer for details, especially concerning the construction of the 
intrinsic state and the symmetries it possesses.  For a given IBM 
Hamiltonian $H$ and intrinsic state $| N, {\bf x} \rangle$, the 
SCC around the $x$-axis is described by 
\begin{equation} 
\delta \langle N, {\bf x} | H - \omega \hat L_x | N, {\bf x} \rangle =0,
\label{ce}
\end{equation}
where $\hat L_x$ is the $x$-component of the angular momentum 
operator, $N$ is the boson number and ${\bf x}$ are the variational 
parameters to be determined from the extremum condition.  For 
convenience, we consider a general formulation of the IBM which will 
be tailored to specific cases later.  Thus, we introduce the boson 
creation and annihilation operators $b^\dagger_{l\mu}, b_{l\mu}$ with 
$l=0,2,4, \ldots$ where $b_0=s$, $b_2=d$, $b_4=g$, etc.  The intrinsic 
state is given by a condensate of intrinsic bosons as
\begin{equation}
| N, {\bf x} \rangle =(N!)^{-1/2}(b^\dagger)^N|0\rangle,\quad
b^\dagger=\sum_{l} \sum_{m=-l}^l x_{lm} b^\dagger_{lm},
\label{is}
\end{equation}
where $x_{lm}$ are the normalized boson mean fields, i.e.  ${\bf 
x}\cdot{\bf x}=1$.  Due to the symmetries in the cranked system 
\cite{sch85}, $x_{lm}$ are real and $x_{l-m}=x_{lm}$.  Note that the 
odd-$m$ components of $x_{lm}$ vanish in the static limit 
($\omega=0$), but not in general.  Thus, invoking the normalization 
condition (e.g.~setting $x_{00}=1$), there are 3 independent 
variational parameters in the $sd$-IBM ($x_{20}, x_{21}, x_{22}$) and 
5 more in the $sdg$-IBM ($x_{40}, \ldots, x_{44}$).  For the IBM 
Hamiltonian, we use the generic multipole form
\begin{equation}
H = \sum_l \varepsilon_l \hat n_l - \sum_{k=0}^{2l_{\max}}
\kappa_k T^{(k)}\cdot T^{(k)}, 
\label{ham}
\end{equation}
where the boson number and multipole operators are given by
\begin{equation}
\hat n_l = \sum_\mu b^\dagger_{l\mu} b_{l\mu}, \quad
T^{(k)}_\mu = \sum_{jl} t_{kjl}[b_j^\dagger \tilde b_l]^{(k)}_\mu.
\label{ops}
\end{equation}
In particular, the angular momentum operators are
\begin{equation}
\hat L_x = {-1\over \sqrt{2}} (\hat L_{+1}-\hat L_{-1}), \quad 
\hat L_\mu= \sum_{l} \sqrt{l(l+1)(2l+1)/3}\ [b_l^\dagger \tilde b_l]^{(1)}_\mu.
\label{lx}
\end{equation}
The parameters of the Hamiltonian consists of the single boson 
energies, $\varepsilon_l$, the multipole coupling strengths, $\kappa_k$, 
and the multipole parameters, $t_{kjl}$.  The hermiticity condition 
on multipole operators requires that $t_{kjl}=t_{klj}$.

The expectation values in Eq.~(\ref{ham}) can be evaluated in a 
straightforward manner using boson calculus techniques.  For the 
one-body operators, one simply obtains
\begin{equation}
\langle N, {\bf x} | \hat n_l | N, {\bf x} \rangle =
N \sum_m {x_{lm}^2 \over {\bf x}\cdot{\bf x}}.
\label{1b}
\end{equation}
Calculations for the multipole interactions are somewhat more 
involved but the expectation values can again be reduced to a compact 
form
\begin{equation}
\langle N, {\bf x} | T^{(k)}\cdot T^{(k)} | N, {\bf x} \rangle 
= N(N-1) \sum_{\mu=-k}^{k} (-1)^\mu A_{k\mu} A_{k-\mu} 
+ N \sum_{lm} \varepsilon'_l {x_{lm}^2 \over {\bf x}\cdot{\bf x}}.
\label{2b}
\end{equation}
Here $A_{k\mu}$ corresponds to the expectation value of the operator 
$T^{(k)}_\mu$ in a single boson state and is given by
\begin{equation}
A_{k\mu} = \sum_{jnlm} (-1)^m \langle jn l m |k \mu\rangle 
t_{kjl} {x_{jn} x_{lm} \over {\bf x}\cdot{\bf x}}.
\label{ak}
\end{equation}
Using the symmetries imposed on the mean fields and the hermiticity 
condition, it can be shown that $A_{k\mu}$ vanishes if $k+\mu$ is odd, 
and furthermore $A_{k-\mu}=(-1)^k A_{k\mu}$.  Thus, for even 
multipoles, only the positive, even $\mu$ values contribute to the sum 
in Eq.~(\ref{2b}).  The second term in Eq.~(\ref{2b}) arises from the 
normal ordering of the boson operators and corresponds to an effective 
one-body term with boson energies
\begin{equation}
\varepsilon'_l = {2k+1 \over 2l+1} \sum_{j}  t_{kjl}^2. 
\end{equation}
Finally, the expectation value of $\hat L_x$ is 
\begin{equation}
\langle N, {\bf x} | \hat L_x | N, {\bf x} \rangle =
-N \sum_{lmn} (-1)^m \langle ln l -m |1 1\rangle \sqrt{2l(l+1)(2l+1)/3}\ 
{x_{ln} x_{lm} \over {\bf x}\cdot{\bf x}}.
\label{lxe}
\end{equation}

For a given angular frequency $\omega$, the cranking equation 
(\ref{ce}) is solved numerically by minimizing the cranking expression 
$E({\bf x}, \omega) =\langle H - \omega L_x \rangle$ with respect to 
the mean fields {\bf x}.  The dynamical MOI is calculated from the 
derivative of $\langle \hat L_x \rangle$, Eq.~(\ref{lxe}) as 
\cite{bm81}
\begin{equation}
{\cal J}^{(2)} = {d \langle \hat L_x \rangle\over d\omega},
\label{j2cr}
\end{equation}
which is obtained under the assumption $\langle L_x \rangle\simeq L$.  
For the exact energy levels $E(L)$, dynamical MOI and the 
corresponding rotational frequencies are calculated from \cite{bm81}
\begin{eqnarray}
 &=& \hbar^2 \left[{d^2 E(L)\over dL^2} \right]^{-1} \simeq 
{4\hbar^2 \over E(L+2)-2E(L)+E(L-2)}, \nonumber \\
\omega &=& {1\over 4\hbar} [E(L+2)-E(L-2)].
\label{j2ex}
\end{eqnarray}
Evolution of triaxiality with rotation is studied using the 
following expressions for the collective shape variable $\gamma$ 
\begin{eqnarray}
\tan\gamma &=& \sqrt{2}\; {\langle Q_2 \rangle \over \langle Q_0 
\rangle},  
\label{gam1} \\
\cos 3\gamma &=& -\sqrt{7 \over 2}\; {\langle Q\cdot Q\cdot Q \rangle \over 
\langle Q\cdot Q \rangle^{3/2}} .
\label{gam2}
\end{eqnarray}
The one-body expectation values of the quadrupole operator in 
Eq.~(\ref{gam1}) follow from Eq.~(\ref{ak}) as $\langle Q_\mu 
\rangle=NA_{2\mu}$.  The two-body expectation value in Eq.~(\ref{gam2})
is given in Eq.~(\ref{2b}) and the three-body part is derived in 
Appendix A.  Eq.~(\ref{gam1}) gives the average value for $\gamma$
while  Eq.~(\ref{gam2}) probes its fluctuations from this average.
Thus the information content of the two expressions for $\gamma$ are 
different and they compliment each other.

\section{Cranking in the \lowercase{$sd$}-IBM}
In this section, we perform cranking calculations for a variety of 
$sd$-IBM Hamiltonians and compare the results for the dynamical MOI 
and $\gamma$ with the exact ones obtained from dynamical symmetries 
(SU(3) and O(6)), and from numerical diagonalization.  For this 
purpose, we use a simple Hamiltonian with a quadrupole interaction and 
$d$-boson energy
\begin{equation}
H = - \kappa Q\cdot Q + \varepsilon \hat n_d.
\label{sdham}
\end{equation}
Following the convention, we denote the parameters of the quadrupole 
operator by $t_{202}=1$ and $t_{222}=\chi$.  The SU(3) limit is 
obtained when $\varepsilon=0$ and $\chi=-\sqrt{7}/2$, and the O(6) 
limit when $\varepsilon=\chi=0$.  To test more realistic 
Hamiltonians, we use the middle value for $\chi$ ($-\sqrt{7}/4$), 
which leads to a better description of electromagnetic properties 
\cite{cas88}.  Inclusion of the $d$-boson energy with 
$\varepsilon=1.5 N\kappa$ improves, in addition, the moment of 
inertia systematics \cite{kuy95b}.  The exact energy eigenvalues 
$E(L)$ are obtained from the Casimir operators in the case of the 
dynamical symmetries, and by numerical diagonalization of the 
Hamiltonian in the latter two cases.

For convenience in the variational problem, we set $x_{00}=1$ and drop 
the subscript 2 from the quadrupole mean fields, i.e.~$x_{2m}=x_m$.  
The normalization is then given by 
\begin{equation}
{\cal N} = 1 + x_0^2 + 2x_1^2 + 2x_2^2.
\label{normsd}
\end{equation}
The expectation value of the Hamiltonian (\ref{sdham}), 
$E({\bf x}) = \langle N, {\bf x} | H | N, {\bf x} \rangle$,
can be written from Eqs.~(\ref{1b}-\ref{2b}) as
\begin{equation}
E({\bf x}) = -\kappa \left[ N(N-1) \left(A_{20}^2 + 2A_{22}^2\right) 
+5\langle \hat n_s \rangle+(\chi^2+1) \langle \hat n_d \rangle \right] 
+ \varepsilon \langle \hat n_d \rangle. 
\label{sde}
\end{equation}
Here the quadratic forms $A_{2\mu}$ follow from Eq.~(\ref{ak}) as
\begin{eqnarray}
A_{20}&=& {1 \over \cal N} \left[ 2x_0 - \sqrt{2/7}\chi\left(x_0^2 
+ x_1^2 - 2x_2^2\right) \right], \nonumber \\
A_{22}&=& {1 \over \cal N} \left[ 2x_2 + \sqrt{2/7}\chi\left (2x_0 x_2 
+ \sqrt{3/2}\, x_1^2\right) \right], 
\label{a2sd}
\end{eqnarray}
and the expectation values of $\hat n_s$ and $\hat n_d$ are
\begin{equation}
\langle \hat n_s \rangle = {N \over \cal N}, \quad 
\langle \hat n_d \rangle = {N \over \cal N} \left( x_0^2 + 2x_1^2 + 
2x_2^2\right).
\label{1bsd}
\end{equation}
From Eq.~(\ref{lxe}), the cranking term is given by
\begin{equation}
\langle \hat L_x  \rangle =
{2N \over \cal N} x_1 \left( \sqrt{6} x_0 + 2x_2 \right).
\label{sdl}
\end{equation}

In Fig.~\ref{fig1}, we compare the cranking results for the dynamical 
MOI obtained from Eq.~(\ref{j2cr}) with those obtained from 
Eq.~(\ref{j2ex}) using the exact energy levels for $N=10$ bosons.  In 
the SU(3) and O(6) limits, the MOI is constant and given by ${\cal 
J}^{(2)}_{\rm ex}= 4/3\kappa = 200/3$ and 
$2/\kappa=100~\hbar^2$/MeV, respectively.  Note that both results 
are independent of $N$, and the constancy of MOI simply follows from 
the fact that $E(L)$ is quadratic in $L$ in both cases.  The SCC leads 
to a constant MOI in the SU(3) limit, as expected from a rigid rotor, 
but deviates from the exact result by about 5\%.
The O(6) limit corresponds to a $\gamma$-unstable rotor for which 
cranking is not expected to work well.  Not surprisingly, comparison 
of the cranking results for ${\cal J}^{(2)}$ in Fig.~\ref{fig1} 
indicate a maximum deviation ($\sim$~10\%) from the exact result, as 
well as a small frequency dependence in MOI. We will comment on these 
deviations further when discussing the $N$ dependence of the results 
below.  The case with $\chi=-\sqrt{7}/4$ falls in between the two 
dynamical symmetry results.  The dynamical MOI increases with $\omega$ 
as in a typical deformed nucleus (Fig.~\ref{fig1}), though the overall
magnitude is still too large.  The deviation between the cranking and 
the exact results in this case is comparable to that in the SU(3) 
limit, with the error getting smaller at higher frequencies.  When a 
$d$-boson energy with $\varepsilon=1.5 N\kappa$ is included in the 
Hamiltonian, which fits the observed range of MOI data better, the 
agreement between the cranking and exact results improve markedly, 
especially at lower frequencies.

The SCC is a semi-classical theory and it would be exact in the 
classical limit when $N \to \infty$.  Thus to understand the nature of 
discrepancies seen in Fig.~\ref{fig1} better, we need to study the $N$ 
dependence of the results.  In Fig.~\ref{fig2}, we show the SCC 
results for ${\cal J}^{(2)}$ as a function of $N$ in the SU(3) and 
O(6) limits.  Since the MOI is not constant in the O(6) limit, we have 
taken the value of ${\cal J}^{(2)}$ at the middle-frequency for each 
$N$.  The curves that trace the SCC results are obtained from
\begin{equation}
{\cal J}^{(2)}_{\rm cr} = {\cal J}^{(2)}_{\rm ex} \left( 1 +
{1 \over 2N} + {1 \over 7N^2} \right),
\label{line1}
\end{equation}
in the SU(3) limit, and from
\begin{equation}
{\cal J}^{(2)}_{\rm cr} = {\cal J}^{(2)}_{\rm ex} \left( 1 -
{1 \over N} + {1 \over 2N^2} \right),
\label{line2}
\end{equation}
in the O(6) limit.  The coefficients in Eqs.~(\ref{line1}) and 
(\ref{line2}) are derived from a $1/N$ expansion of the SCC equation 
(\ref{ce}).  It is clear from Fig.~\ref{fig2} and Eqs.~(\ref{line1}, 
\ref{line2}) that the SCC calculation of ${\cal J}^{(2)}$ is correct 
to leading order but fails in higher orders in $1/N$ by generating 
spurious correction terms.  A comparison of Eq.~(\ref{sde}) with the 
accurate angular-momentum-projected result \cite{kuy88} confirms that 
the incorrect treatment of the $1/N$ terms in the SCC is the cause of 
the discrepancy.  It is interesting to note that an agreement is 
obtained if one uses the Casimir operator of the SU(3) given by $C_2 = 
Q\cdot Q +(3/8) L\cdot L$, instead of just $Q\cdot Q$.  In this case, 
the incorrect $1/N$ contribution from the $Q\cdot Q$ interaction is 
exactly canceled by the spurious contribution from the $L\cdot L$ 
term.  Another interesting observation is that the best agreement of 
SCC with the exact results is achieved in the realistic case with 
$\chi=-\sqrt{7}/4$ and $\varepsilon=1.5 N\kappa$.  This happens 
because the $1/N$ term is now dominated by the leading contribution 
from $\langle \hat n_d \rangle$, which is correctly treated in the 
SCC. The errors from $\langle \hat n_d \rangle$ contribute at the 
$1/N^2$ level, and these become apparent only at the high-rotational 
frequencies as is apparent in Fig.~\ref{fig1}.

We next study the effects of triaxiality in SCC. It is well known that 
the energy surface of an $sd$-IBM Hamiltonian with one- and two-body 
interactions has an axially symmetric minima in the deformed phase 
\cite{gin80}.  An exception occurs in the O(6) limit, where a 
$\gamma$-unstable shape develops.  The SCC calculations in the O(6) 
limit give $30^{\rm o}$ for both the average $\gamma$ (\ref{gam1}) and 
its fluctuations (\ref{gam2}) at all frequencies.  Thus the SCC 
results are consistent with the description of the O(6) limit using a 
triaxial intrinsic state with $\gamma=30^{\rm o}$, which has a very 
shallow minimum in the $\gamma$ direction \cite{os87}.  Frequency 
independence of the results corroborates with the fact that the O(6) 
states remain $\gamma$-unstable at all spins.  In Fig.~\ref{fig3}, we 
show the evolution of the triaxiality angle $\gamma$ with the cranking 
frequency for the remaining three cases discussed above for three 
different boson numbers, $N=10$, 15, 20.  Results obtained from both 
Eq.~(\ref{gam1}) (solid line) and (\ref{gam2}) (dashed line) are 
shown.  The ground band in the SU(3) limit can be described exactly by 
angular momentum projection from an axially symmetric intrinsic state, 
therefore, both average $\gamma$ and its fluctuations should vanish at 
{\em all} frequencies.  The SCC results (Fig.~\ref{fig3}, left) start 
with $\gamma=0$ at low $\omega$ but deviate from it systematically 
with increasing frequency.  Although the situation appears to be 
improving with increasing $N$, similar to the case in ${\cal 
J}^{(2)}$, in fact, this is merely due to the extension of the spectrum 
with $N$ ($\omega_{\max}\propto N$).  If one scales out the $N$ 
dependence (i.e., plots $\gamma$ against $\omega/N$), then all the 
curves with different $N$ overlap.  Thus $\gamma$ can be written as a 
power expansion in $\omega/N$ with the leading ($N$ independent) term 
being zero.  This situation is similar to that encountered in the 
study of ${\cal J}^{(2)}$: the SCC gets the leading order term 
correctly ($\gamma=0$) but fails in higher-orders in $1/N$ by 
generating spurious terms in powers of $\omega/N$.  The fact that the 
incorrect $1/N$ terms in $\gamma$ all depend on the cranking frequency 
makes the SCC results increasingly unreliable at higher frequencies.  
The onset of triaxiality observed at around mid-frequency 
($\omega_{\max}/2$) in all curves in Fig.~\ref{fig3} is thus due the 
failure of the SCC and not a genuine feature of the boson system.  We 
remark that despite the sudden increase in the calculated $\gamma$ 
values, ${\cal J}^{(2)}$ remains constant in the SU(3) limit, 
suggesting that triaxiality has a negligible effect on MOI.

We interpret the remaining more realistic cases in the light of the 
SU(3) test.  In the case of $\chi=-\sqrt{7}/4$ (Fig.~\ref{fig3}, 
middle), the average $\gamma$ (solid lines) remains around zero except 
at high frequencies.  Attributing this sudden rise in $\gamma$ to the 
break down of SCC, we see that $\gamma \sim 0$ for all ground-band 
states of a $Q\cdot Q$ Hamiltonian.  The fluctuations (dashed lines), 
on the other hand, are non-zero but get smaller with increasing $N$.  
The values of fluctuations at $\omega=0$ are consistent with those 
obtained from an exact calculation for the ground state \cite{ell86}.  
At higher frequencies there is a small reduction in fluctuations, that 
is, rotations have a slightly stabilizing effect on the $\gamma$ 
motion.  Increasing the boson number also suppresses fluctuations, 
which is due to the energy surface becoming deeper in the $\gamma$ 
direction with larger $N$ values.  The last case with 
$\varepsilon=1.5 N\kappa$ (Fig.~\ref{fig3}, right) has broadly the 
same features.  The one-body term cannot induce triaxiality alone, 
therefore, the small deviation in average $\gamma$ from zero at 
mid-frequencies is presumably due to errors in SCC. There is an 
increase in fluctuations compared to the middle panel, which can be 
explained as due to the energy surface becoming shallower with the 
addition of the one-body term.  These results indicate that rotations 
have a rather limited effect on the shape variables in the $sd$-IBM, 
with average $\gamma$ remaining nearly constant at zero, and its 
fluctuations being slightly reduced from the ground state value at 
higher spins.

\section{Cranking in the \lowercase{$sdg$}-IBM}
In contrast to the $sd$-IBM, where numerical diagonalization is a 
routine task, the $sdg$-IBM already suffers from the large basis 
problem, and exact diagonalization is not possible for $N > 11$, that 
is, for most of the deformed nuclei.  The $1/N$ expansion circumvents 
this problem but because it assumes axial symmetry, one cannot use it 
to address questions on triaxiality.  Here we perform cranking 
calculations in the $sdg$-IBM to study the evolution of shapes with 
rotation without restriction to axial symmetry.  For the Hamiltonian, 
we choose
\begin{equation}
H= - \kappa_2 Q \cdot Q - \kappa_4 T_4 \cdot T_4 
+ \varepsilon_d \hat n_d + \epsilon_g \hat n_g,
\label{hamsdg}
\end{equation}
which has been shown to provide a good representation of data in 
rare-earth and actinide nuclei \cite{san96}.  As in Section III, we 
set $q_{02}=h_{02}=1$ and $x_{00}=1$.  Then the normalization is given by
\begin{equation}
{\cal N} = 1 + x_{20}^2 + 2x_{21}^2 + 2x_{22}^2 
+ x_{40}^2 + 2x_{41}^2 + 2x_{42}^2 + 2x_{43}^2 + 2x_{44}^2.
\label{normsdg}
\end{equation}
The expectation value of the Hamiltonian (\ref{hamsdg}) follows from
Eqs.~(\ref{1b}-\ref{2b}) as
\begin{eqnarray}
E({\bf x}) &=& -\kappa_2 \left[ N(N-1) (A_{20}^2 + 2A_{22}^2) 
+5\langle \hat n_s \rangle + (1+q_{22}^2+q_{24}^2) \langle \hat n_d \rangle 
+ (5/9)(q_{24}^2+q_{44}^2) \langle \hat n_g \rangle  \right]
\nonumber \\
&& -\kappa_4 \Bigl[ N(N-1) (A_{40}^2 + 2A_{42}^2 + 2A_{44}^2)
+9\langle \hat n_s \rangle + (9/5)(h_{22}^2+h_{24}^2) \langle \hat n_d \rangle 
+ (1+h_{24}^2+h_{44}^2) \langle \hat n_g \rangle \Bigr] \nonumber \\
&& + \varepsilon_d \langle \hat n_d \rangle
+ \varepsilon_g \langle \hat n_g \rangle. 
\label{sdge}
\end{eqnarray}
Here the quadrupole quadratic forms $A_{2\mu}$ are given by
\begin{eqnarray}
A_{20}&=& {1 \over \cal N} \left[ 2x_{20} 
- \sqrt{2/7} q_{22} \left(x_{20}^2 + x_{21}^2 - 2x_{22}^2\right) 
+2\sqrt{2/21} q_{24} \left(\sqrt{3}x_{20}x_{40} + \sqrt{10}x_{21}x_{41} 
\right. \right. \nonumber \\
&&\hskip .8cm  + \sqrt{5}x_{22}x_{42} \Bigr)
-(1/3\sqrt{77}) q_{44} \left(10x_{40}^2 + 17x_{41}^2 
+ 8x_{42}^2-7x_{43}^2 - 28x_{44}^2\right) \biggr], \nonumber \\
A_{22}&=& {1 \over \cal N} \biggl[ 2x_{22} 
+ \sqrt{2/7} q_{22} \left( 2x_{20}x_{22} + \sqrt{3/2}x_{21}^2\right) 
\nonumber \\
&&\hskip .8cm +(2/3\sqrt{14}) q_{24} \left(x_{22}x_{40} + \sqrt{5}x_{21}x_{41} 
+ \sqrt{15}x_{20}x_{42}+\sqrt{35}x_{21}x_{43}+\sqrt{70}x_{22}x_{44}\right)
\nonumber \\
&&\hskip .8cm + \sqrt{2/33} q_{44} \left(2x_{42}x_{44} + 3x_{41}x_{43} 
+3\sqrt{10/7}x_{40}x_{42} + (5/\sqrt{7}) x_{41}^2 \right) \biggr], 
\label{a2sdg}
\end{eqnarray}
and the hexadecapole quadratic forms $A_{4\mu}$ by
\begin{eqnarray}
A_{40}&=& {1 \over \cal N} \left[ 2x_{40} 
+ \sqrt{2/35} h_{22} \left(3x_{20}^2 -4 x_{21}^2 + x_{22}^2\right) 
-(2/\sqrt{77}) h_{24} \left(2\sqrt{5}x_{20}x_{40} + \sqrt{6}x_{21}x_{41} 
\right. \right. \nonumber \\
&&\hskip .8cm  - 6\sqrt{3}x_{22}x_{42} \Bigr)
+ \sqrt{2/1001} h_{44} \left(9x_{40}^2 + 9x_{41}^2 
-11x_{42}^2 -21x_{43}^2 +14x_{44}^2\right) \biggr], \nonumber \\
A_{42}&=& {1 \over \cal N} \biggl[ 2x_{42} 
+ (1/\sqrt{7}) h_{22} \left(\sqrt{6}x_{20}x_{22} -2x_{21}^2\right) 
+ \sqrt{3/385} h_{24} \nonumber \\
&&\hskip .8cm \times \left(6\sqrt{5} x_{22}x_{40} 
+ 9x_{21}x_{41} -(8/\sqrt{3})x_{20}x_{42} -5\sqrt{7}x_{21}x_{43}
+2\sqrt{14}x_{22}x_{44}\right)
\nonumber \\
&&\hskip .8cm + \sqrt{5/1001} h_{44} \left(6\sqrt{7}x_{42}x_{44} + 
2\sqrt{7}x_{41}x_{43} -11x_{40}x_{42} -6x_{41}^2 \right) \biggr], 
\nonumber \\
A_{44}&=& {1 \over \cal N} \biggl[ 2x_{44} + h_{22} x_{22}^2  
+(2/\sqrt{55}) h_{24} \left( \sqrt{6} x_{22}x_{42} 
+ \sqrt{21}x_{21}x_{43} +2\sqrt{7}x_{20}x_{44} \right)
\nonumber \\
&&\hskip .8cm + (1/\sqrt{143}) h_{44} \left( 2\sqrt{14}x_{40}x_{44}  
+2\sqrt{35}x_{41}x_{43} +3\sqrt{5}x_{42}^2 \right) \biggr].
\label{a4sdg}
\end{eqnarray}
The one-body expectation values in Eq.~(\ref{sdge}) are given by
Eq.~(ref{1bsd}) and 
\begin{equation}
\langle \hat n_g \rangle = {N \over \cal N} 
\left(x_{40}^2 + 2x_{41}^2 + 2x_{42}^2 + 2x_{43}^2 + 2x_{44}^2\right). 
\end{equation}
Finally, from Eq.~(\ref{lxe}) the cranking term is
\begin{equation}
\langle \hat L_x  \rangle =
{2N \over \cal N} \left[ x_{21} \left( \sqrt{6} x_{20} + 2x_{22} \right)
+ x_{41} \left( 2\sqrt{5} x_{40} + 3\sqrt{2}x_{42} \right)
+ x_{43} \left( \sqrt{14} x_{42} + 2\sqrt{2}x_{44} \right)\right].
\label{sdgl}
\end{equation}

In view of the numerous parameters in the $sdg$-IBM, we limit our 
discussion to a realistic range tailored to data \cite{san96}. 
Accordingly, the quadrupole parameters $\{q_{22},q_{24},q_{44}\}$ are 
scaled from their SU(3) values with a single factor $q$ as suggested 
by microscopics \cite{oai78}.  The hexadecapole parameters 
$\{h_{22},h_{24},h_{44}\}$ are determined from those of $q_{jl}$ 
through the commutation condition
\begin{equation}
[\bar{h},\bar{q}]=0, \quad \bar{q}_{jl}=\langle j0l0|20\rangle q_{jl},
\quad \bar{h}_{jl}=\langle j0l0|40\rangle h_{jl},
\end{equation}
which reproduce the available $E4$ data reasonably well \cite{san96}.  
We adapt the realistic $sd$-IBM parameters used in Fig.~\ref{fig1}, 
$\kappa_2=-20$~keV, $q=0.5$ and $\epsilon_d=1.5 N\kappa_2$, and use 
further, $\epsilon_g=4N\kappa_2$ and $\kappa_4/\kappa_2=0$-0.4.  
Energy levels are very well described in the $sdg$-IBM~\cite{san96}, 
therefore we do not dwell on a study of MOI apart from noting that all 
the SCC calculations of ${\cal J}^{(2)}$ exhibit the characteristic 
rise with increasing frequency seen in experimental spectra.  We focus 
instead on the shape question which could not be addressed in the 
$1/N$ expansion approach.  With the parameter set described above, the 
SCC calculations of $\gamma$ in the $sdg$-IBM are very similar to 
those of the $sd$-IBM with the realistic set of parameters 
(Fig.~\ref{fig3}, right).  Because the hexadecapole interaction is a 
relatively novel feature not used in the $sd$-IBM, we briefly comment 
on its effect.  Inclusion of the hexadecapole interaction does not 
cause any deviation from the axial shape but depending on its sign, it 
either decreases the fluctuations in $\gamma$ (attractive) or 
increases them (repulsive) by a few degrees.  This is very similar to 
the effect of changing the strength of the quadrupole interaction, 
in that, a larger $\kappa_2$ leads to a deeper energy well in the 
$\gamma$ direction, therefore reduces the fluctuations, and vice 
versa.  In certain parametrizations where the boson energies are 
neglected, it is possible to obtain non-axial shapes in the $sdg$-IBM 
\cite{kuy91}.  However, once a $g$-boson energy of $\epsilon_g\sim 
1$~MeV is included, as demanded by the experimental spectra, such 
deviations from axial symmetry quickly disappear.  In short, the 
congruence found between the $sd$- and $sdg$-IBM results with regard 
to the shape variable $\gamma$ is a direct consequence of the fairly 
high $g$-boson energy, which limits the effect of $g$ bosons to a 
perturbative range at low to medium spins.

\section{Conclusions}
We have tested the self-consistent cranking method using the exact IBM 
results obtained from dynamical symmetries and numerical 
diagonalization.  The SCC results for the dynamic MOI are in good 
agreement with the exact ones, especially in realistic cases where the 
deviation is at most a few percent.  The study of shapes using SCC, on 
the other hand, remains problematic due to the generation of spurious 
frequency-dependent terms in shape variables, which become dominant at 
high frequencies.  Nevertheless, SCC can give reliable information on 
evolution of shapes in the low to mid frequency range.  Our study of 
shapes using SCC indicates that both the $sd$- and $sdg$-IBM with one- 
and two-body interactions retain the axial shape ($\gamma=0$) with 
fluctuations around $10^{\rm o}$, as long as the model parameters are 
restricted to a realistic range that reproduce the data.  The only 
frequency dependence in shape variables is observed in the 
fluctuations in $\gamma$ which is slightly reduced with increasing 
frequency.  These results suggest that if the triaxial effects are 
important and need to be included in the IBM, then the remedy should 
be sought in three-body interactions rather than $g$ bosons.

\appendix
\section{Three-body operators}
Here we derive the expectation value of the three-body quadrupole 
interaction in Eq.~(\ref{gam2}).  The scalar product of the three 
quadrupole operators is defined as
\begin{equation}
Q\cdot Q\cdot Q = [Q Q]^{(2)} \cdot Q.
\end{equation}
The expectation value of this three-body term in the condensate state 
Eq.~(\ref{is}) is given by
\begin{equation}
\langle Q\cdot Q\cdot Q \rangle = {1\over N!} \sum_{\mu_1\mu_2\mu} 
(-1)^{\mu} \langle 2 \mu_1 2 \mu_2 |2 \mu\rangle
\langle 0| b^N  Q_{\mu_1} Q_{\mu_2} Q_{-\mu} (b^\dagger)^N |0\rangle.
\end{equation}
Writing the quadrupole operators explicitly and commuting all the 
boson creation operators to the left, one obtains a three-body term, 3 
two-body terms and a one-body term.  Using boson calculus, the matrix 
elements of these normal ordered operators can be evaluated in a 
straightforward manner with the result
\begin{eqnarray}
\langle Q\cdot Q\cdot Q \rangle &=&  \sum_{\mu_1\mu_2\mu} 
(-1)^{\mu} \langle 2 \mu_1 2 \mu_2 |2 \mu\rangle 
\sum_{j_1j_2j_3 l_1l_2l_3\atop m_1m_2m_3 n_1n_2n_3} 
(-1)^{n_1+n_2+n_3} q_{j_1l_1} q_{j_2l_2} q_{j_3l_3}\nonumber \\
&& \times \langle j_1 m_1 l_1 n_1 |2 \mu_1\rangle
\langle j_2 m_2 l_2 n_2 |2 \mu_2\rangle
\langle j_3 m_3 l_3 n_3 |2 -\mu\rangle \nonumber \\
&& \times \{ N(N-1)(N-2) x_{j_1 m_1} x_{j_2 m_2} x_{j_3 m_3}
x_{l_1 -n_1} x_{l_2 -n_2} x_{l_3 -n_3} \nonumber \\
&& + N(N-1) [x_{j_1 m_1} x_{j_2 m_2} x_{l_2 -n_2} x_{l_3 -n_3} 
\delta_{j_3l_1} \delta_{m_3 -n_1} + x_{j_1 m_1} x_{j_2 m_2} \nonumber \\
&& \quad \times x_{l_1 -n_1} x_{l_3 -n_3} 
\delta_{j_3l_2} \delta_{m_3 -n_2} 
+ x_{j_1 m_1} x_{j_3 m_3} x_{l_2 -n_2} x_{l_3 -n_3} 
\delta_{j_2l_1} \delta_{m_2 -n_1}] \nonumber \\
&& + N x_{j_1 m_1} x_{l_3 -n_3} 
\delta_{j_2l_1} \delta_{m_2 -n_1} \delta_{j_3l_2} \delta_{m_3 -n_2} \}.
\label{3bme1}
\end{eqnarray}
The three-body part (the first term) is seen to involve a triple 
product of the one-body expectation values $A_{2\mu}$ defined in 
Eq.~(\ref{ak}).  In the effective two-body terms, three C-G 
coefficients can be summed to yield a $6-j$ symbol and a C-G 
coefficient.  After some manipulations of the summation indices, all 
3 terms can be shown to be equivalent.  Finally, the sum of the 
four C-G coefficients in the effective one-body term gives a $6-j$ 
symbol.  Thus, Eq.~(\ref{3bme1}) can be written compactly as
\begin{eqnarray}
\langle Q\cdot Q\cdot Q \rangle &=&  N(N-1)(N-2) \sum_{\mu_1\mu_2\mu} 
(-1)^{\mu} \langle 2 \mu_1 2 \mu_2 |2 \mu\rangle A_{2\mu_1} A_{2\mu_2} 
A_{2-\mu} \nonumber \\
&&  + 3N(N-1) \sum_{\mu} (-1)^{\mu} \tilde A_{2\mu} A_{2-\mu}
+ N \sum_{lm} \tilde \varepsilon_l x_{lm}^2,
\label{3bme2}
\end{eqnarray}
where we have introduced
\begin{eqnarray}
\tilde q_{jl} &=& 5 \sum_i (-1)^{i+j+l} \left\{ \begin{array}{ccc} 
j&l&2 \\ 2&2&i \end{array} \right\}
q_{ji} q_{il}, \nonumber \\
\tilde \varepsilon_l &=& {25 \over 2l+1} \sum_{ij} 
\left\{ \begin{array}{ccc} l&i&2 \\ 2&2&j \end{array} \right\} 
q_{li} q_{ij} q_{jl},
\label{tilde}
\end{eqnarray}
and $\tilde A_{2\mu}$ is obtained from $A_{2\mu}$ by replacing $q_{jl} 
\to \tilde q_{jl}$ in Eq.~(\ref{ak}).  Note that symmetry of $q_{jl}$ 
is retained, i.e., $\tilde q_{jl}=\tilde q_{lj}$.  Using the fact that 
$A_{21}=A_{2-1}=0$ and $A_{22}=A_{2-2}$, the sums in Eq.~(\ref{3bme2}) 
involving $A_{2\mu}$ can be carried out to yield
\begin{eqnarray}
\langle Q\cdot Q\cdot Q \rangle &=&  N(N-1)(N-2) \sqrt{2/7}
(-A_{20}^2 + 6A_{22}^2) A_{20} 
\nonumber \\
&&  + 3N(N-1)  (\tilde A_{20} A_{20} + 2\tilde A_{22} A_{22}
+ N \sum_{lm} \tilde \varepsilon_l x_{lm}^2,
\label{3bme3}
\end{eqnarray}

In the case of the $sd$-IBM, $A_{20}$ and $A_{22}$ are given in 
Eq.~(\ref{a2sd}).  The contracted parameters in Eq.~(\ref{tilde}) 
become
\begin{eqnarray}
&&\tilde q_{02} = \chi, \quad \tilde q_{22} = 
1-3\chi^2/14, \nonumber \\
&& \tilde \varepsilon_0 = 5\chi, \quad \tilde \varepsilon_2 = 2\chi - 
3\chi^3/14.
\label{tsd}
\end{eqnarray}
Using these values of $\tilde q_{jl}$, one obtains for $\tilde A_{20}$ 
and $\tilde A_{22}$
\begin{eqnarray}
\tilde A_{20}&=& {1 \over \cal N} \left[ 2\chi x_0 - 
\sqrt{2/7}(1-3\chi^2/14) (x_0^2 + x_1^2 - 2x_2^2) \right], \nonumber \\
\tilde A_{22}&=& {1 \over \cal N} \left[ 2\chi x_2 + \sqrt{2/7}(1-3\chi^2/14) 
(2x_0 x_2 + \sqrt{3/2}\, x_1^2) \right], 
\label{ta2sd}
\end{eqnarray}

In the $sdg$-IBM, these parameters are given by 
\begin{eqnarray}
&&\tilde q_{02} = q_{22}, \quad 
\tilde q_{22} = 1-{3\over 14} q_{22}^2 + {2\over 7} q_{24}^2, \quad 
\tilde q_{24} = {2\over 7} q_{22}q_{24} - {5\sqrt{22}\over 42} q_{24} 
q_{44}, \nonumber \\
&&\tilde q_{44} = {5\sqrt{22}\over 42} q_{24}^2 - {65\over 42\sqrt{22}} 
q_{44}^2, \nonumber \\
&& \tilde \varepsilon_0 = 5q_{22}, \quad 
\tilde \varepsilon_2 = 2 q_{22}- {3\over 14} q_{22}^3 + {4\over 7} 
 q_{22}q_{24}^2 + {5\sqrt{22}\over 42} q_{24}^2q_{44}, \nonumber \\ 
&&\tilde \varepsilon_4 = {10\over 63} q_{22}q_{24}^2 + {25 
\sqrt{22}\over 189} q_{24}^2 q_{44} - {325\over 378\sqrt{22}} q_{44}^3.
\label{tsdg}
\end{eqnarray}

\begin{figure}
\caption{Comparison of the cranking calculations of the dynamical 
moment of inertia in the $sd$-IBM (lines) with the exact results 
obtained from the dynamical symmetries SU(3) and O(6) (filled 
circles), and numerical diagonalization ($\chi=-\sqrt{7}/4$ with 
$\varepsilon=0$ and $\varepsilon=1.5 N\kappa$) (open circles).  In 
all cases, $N=10$ and $\kappa=-20$~keV are used.}
\label{fig1}
\end{figure}

\begin{figure}
\caption{Boson number dependence of the cranking dynamical moment of 
inertia in the SU(3) and O(6) limits of the $sd$-IBM (circles).  The 
lines that trace the circles are obtained from Eq.~(\ref{line1}) in 
the SU(3) case and from Eq.~(\ref{line2}) in the O(6) case.}
\label{fig2}
\end{figure}

\begin{figure}
\caption{Cranking calculations of the triaxiality angle $\gamma$ in 
the $sd$-IBM for the three cases shown in Fig.~1 for $N=10,$ 15, 20.  
The solid lines indicate average $\gamma$ obtained from 
Eq.~(\ref{gam1}) and the dashed lines correspond to its fluctuations 
given by Eq.~(\ref{gam2}).}
\label{fig3}
\end{figure}

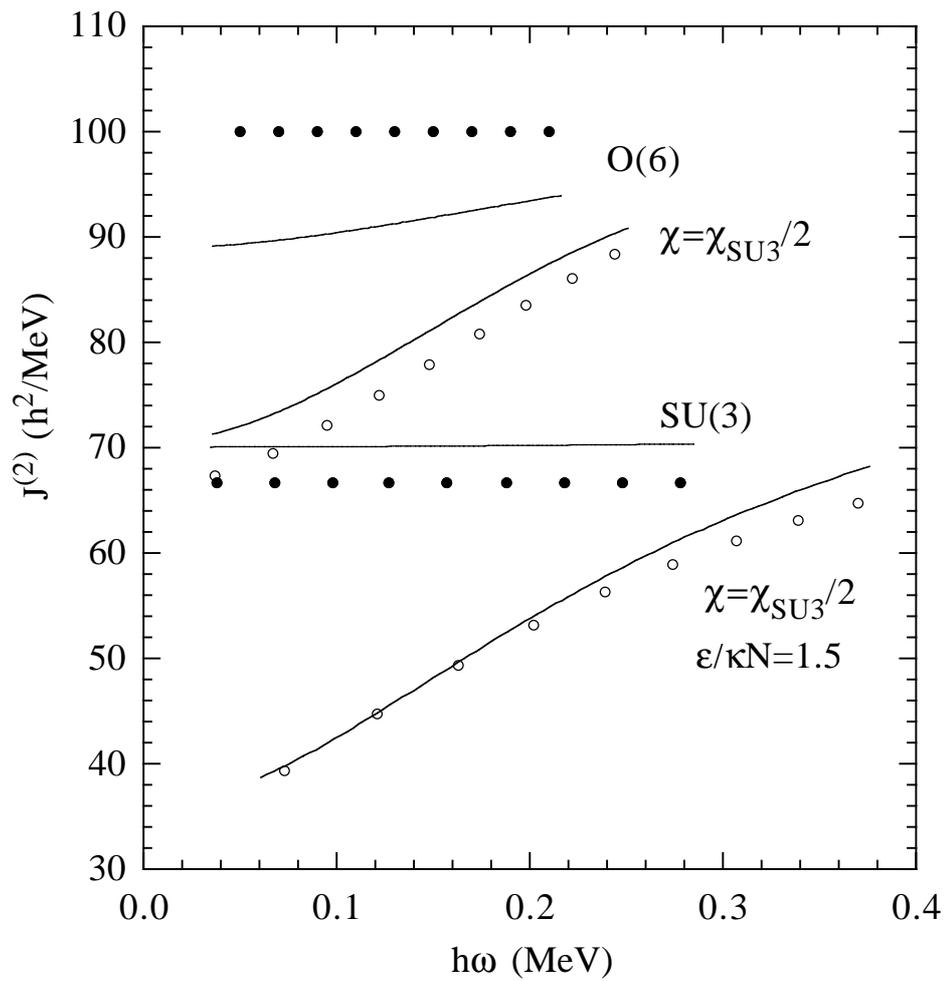
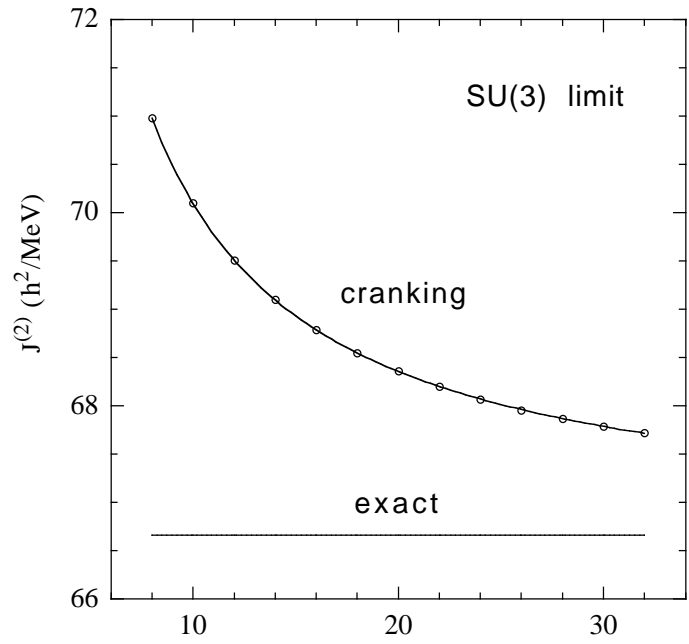
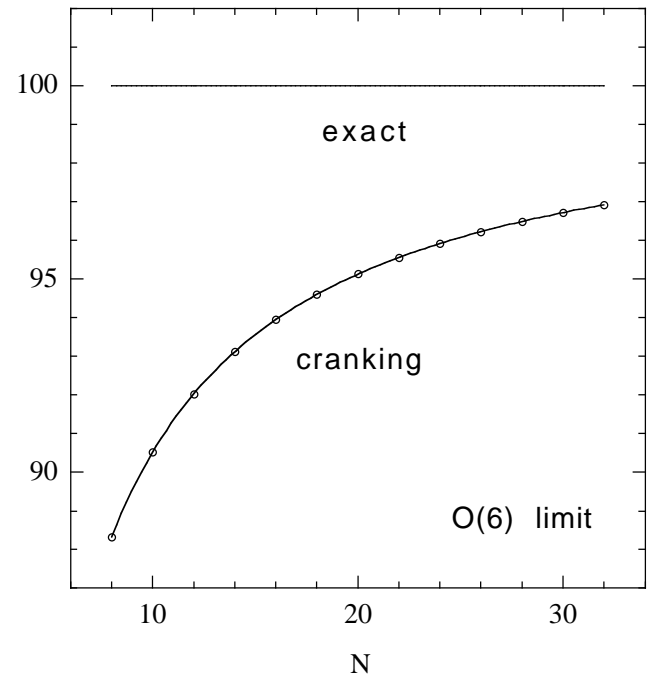
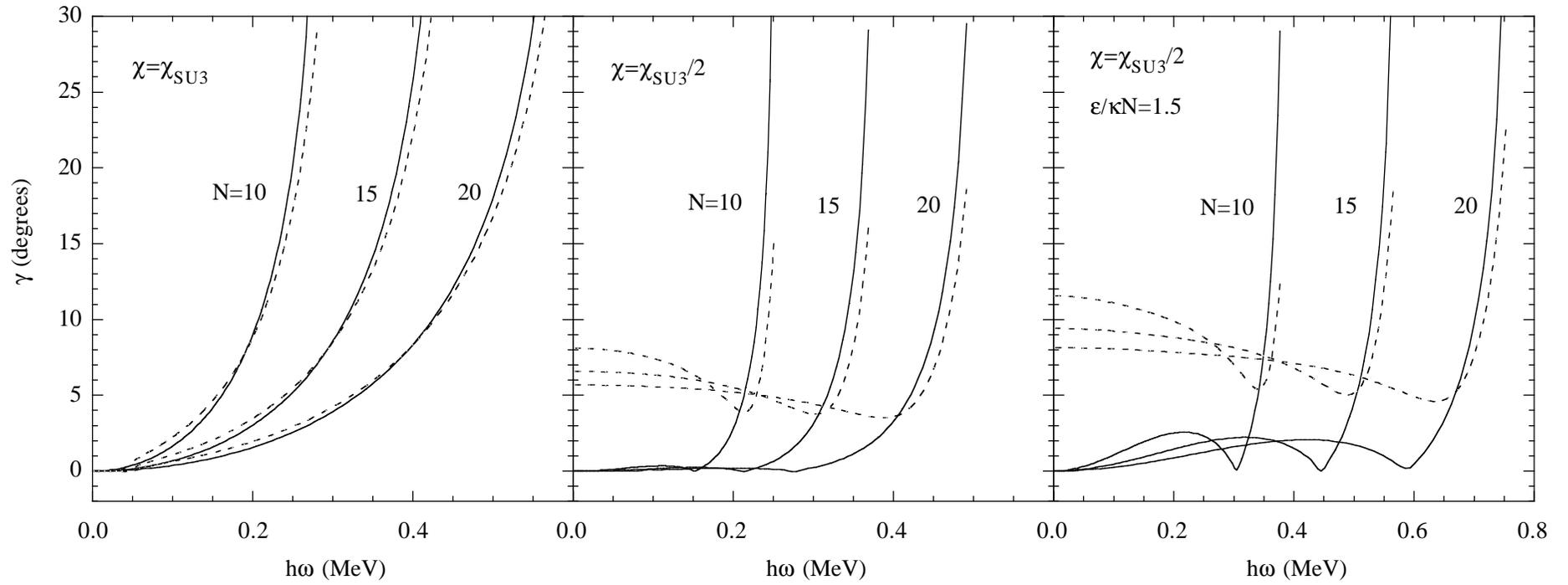
\end{document}